\documentclass[10pt,conference]{IEEEtran}

\usepackage{comment}

\usepackage{graphicx}   
\usepackage{url}        
\usepackage{amssymb}
\usepackage{amsmath}    
\usepackage{hyperref}
\usepackage[export]{adjustbox}
\usepackage{subcaption}
\usepackage{fancyhdr}

\usepackage{physics}
\usepackage{quantikz}
\usepackage{fancyhdr}

\fancypagestyle{specialfooter}{%
\fancyhf{}

\fancyfoot[R]{ \noindent\fbox{%
\parbox{\textwidth}{%
{\footnotesize This work has been submitted to the IEEE for possible publication.
Copyright may be transferred without notice, after which this version may no longer be accessible.}
}
}}
}

\begin{document}
\bstctlcite{IEEEexample:BSTcontrol}

\title{Understanding the effects of data encoding on quantum-classical convolutional neural networks}

\author{
\IEEEauthorblockN{Maureen Monnet\IEEEauthorrefmark{1}, Nermine Chaabani\IEEEauthorrefmark{1}, Theodora-Augustina Dr\u{a}gan\IEEEauthorrefmark{1}, Balthasar Schachtner\IEEEauthorrefmark{2}, Jeanette Miriam Lorenz\IEEEauthorrefmark{1}}
\IEEEauthorblockA{\IEEEauthorrefmark{1}Fraunhofer Institute for Cognitive Systems IKS,  Munich, Germany}
\IEEEauthorblockA{\IEEEauthorrefmark{2}LMU University Hopsital, Munich, Germany}
\{maureen.monnet, theodora-augustina, jeanette.miriam.lorenz\}@iks.fraunhofer.de \\ balthasar.schachtner@med.lmu.de
}
\maketitle\thispagestyle{specialfooter} 

\begin{abstract}

Quantum machine learning was recently applied to various applications and leads to results that are comparable or, in certain instances, superior to classical methods, in particular when few training data is available. These results warrant a more in-depth examination of when and why improvements can be observed. A key component of quantum-enhanced methods is the data encoding strategy used to embed the classical data into quantum states. However, a clear consensus on the selection of a fitting encoding strategy given a specific use-case has not yet been reached. This work investigates how the data encoding impacts the performance of a quantum-classical convolutional neural network (QCCNN) on two medical imaging datasets. In the pursuit of understanding why one encoding method outperforms another, two directions are explored. Potential correlations between the performance of the quantum-classical architecture and various quantum metrics are first examined. Next, the Fourier series decomposition of the quantum circuits is analyzed, as variational quantum circuits generate Fourier-type sums. We find that while quantum metrics offer limited insights into this problem, the Fourier coefficients analysis appears to provide better clues to understand the effects of data encoding on QCCNNs.

\end{abstract}

\begin{IEEEkeywords}
quantum machine learning, data encoding, quantum convolutional neural networks, medical imaging
\end{IEEEkeywords}

\section{Introduction}

Quantum convolutional neural networks (QCNNs) are a promising architecture in the domain of image classification that is theoretically shown to be beneficial in situations where few training data points are available~\cite{caro_generalization_2022}. This makes them particularly interesting for e.g. medical applications where data availability is limited due to data collection and annotation costs, as well as privacy and security concerns. While QCNNs are originally designed to be fully implemented on quantum computers~\cite{cong_quantum_2019}, the currently available quantum hardware -- commonly referred to as Noisy Intermediate-Scale Quantum (NISQ) computers -- are not yet up to the task. More technological development is required in order to implement robust fully-quantum solutions, such as an increased number of qubits, reduced error rates and improved coherence times.

A more NISQ-friendly approach to QCNNs is achieved by running a hybrid quantum-classical version, where one or several chosen components of the classical convolution neural network (CNN) architecture are replaced by a quantum variant. In the classical case, a CNN typically comprises convolutional layers responsible for extracting characteristic features from images, followed by
fully connected layers employed for the classification. In the quantum-classical case, the convolutional layers stand out as promising candidates for their mapping onto a NISQ quantum computer, as they handle data sequentially in smaller segments, in contrast to the linear layers that process all features simultaneously. In that scenario, the convolutional layer is replaced by a variational quantum circuit (VQC) containing trainable operations, while the remainder of the architecture remains classical. The optimization is then performed on a classical computer using a classical optimizer.

In order to process classical data with a quantum circuit, a crucial step is the data encoding, as it defines the feature space the quantum circuit is able to generate (\cite{schuld_quantum_2019,havlicek_supervised_2019}). The choice of a suitable encoding strategy for a given dataset is therefore a non-trivial endeavour, with literature displaying ongoing efforts to better understand its impact on the performance of the entire algorithm \cite{mattern_variational_2021}.
Further techniques related to data encoding, such as data-reuploading, have demonstrated the potential to enhance performance significantly \cite{perez-salinas_data_2020}. In this approach, the encoding scheme is repeated between the different variational layers.

In this work, we evaluate the performance on two medical imaging datasets of QCCNNs with four different encoding strategies, namely the angle encoding around the X- and Y-axis, the higher-order, and the amplitude encoding. The studied datasets include 2D ultrasound images of the breast for identifying malignant lesions, alongside abdominal computed tomography (CT) scans for conducting multi-class classification of body organs. Specifically, we compare the performance of the selected data encoding strategies when they are re-uploaded up to five times. We then strive to understand the differences in performances by employing two distinct approaches: in the first approach, we use quantum metrics proposed in literature to quantify the power of quantum neural networks. In the second approach, we apply the Fourier formalism to the outputs generated by our quantum circuits. The motivation behind pursuing the latter arises from the observation that outputs generated by variational quantum circuits can be expressed in terms of Fourier series \cite{schuld_effect_2021}.

Our contributions are thus as follows:
\begin{itemize}
    \item We analyze the effect of four different data encodings on the performance of a quantum-classical convolutional neural network (QCCNN) for binary and multi-class classifications of ultrasound and CT medical images. We additionally investigate the impact of input scaling and data-reuploading on the results. 
    \item We analyze the performance of all architectures in the light of 3 relevant quantum metrics, namely the expressibility, the entanglement capability and the normalized effective dimension to find possible correlations.
    \item We extend this analysis by applying the Fourier formalism on the outputs produced by the quantum circuits. 
\end{itemize}
The paper is structured as follows. We present the related work in section~\ref{relatedwork}. The background information relevant to the designed QCCNNs and their encodings, the data-reuploading scheme, the three quantum metrics and the Fourier series representation of quantum circuits is introduced in section~\ref{background}. We present the datasets used, the performances of the QCCNNs and their potential correlations with the defined quantum metrics and Fourier coefficients in the results and discussion section~\ref{results}. We conclude in section~\ref{ccl}.

\section{Related Work}
\label{relatedwork}
\subsection{Empirical studies on the performance of QCCNNs}
Fully quantum QCNNs have been studied in several works (\cite{cong_quantum_2019, kerenidis_quantum_2019, hur_quantum_2022, li_quantum_2020, wei_quantum_2021, lu_quantum_2021}). However, with the current limitations posed by NISQ hardware, it is of interest to move only specific components of the architecture to the quantum computer, resulting in a hybrid quantum-classical architecture. The convolutional layer in the classical CNN is substituted with either an untrainable or trainable quantum convolutional layer in (\cite{henderson_quanvolutional_2019, mattern_variational_2021, liu_hybrid_2021, matic_quantum-classical_2022, monnet_pooling_2023}), where good performance is shown on the MNIST dataset in (\cite{henderson_quanvolutional_2019, mattern_variational_2021}). \cite{mattern_variational_2021} particularly focuses on the impact of the encoding scheme and filter size of the quantum convolution, showing that some architectures perform better than others. Building upon this, our work in \cite{matic_quantum-classical_2022} and \cite{monnet_pooling_2023} introduces diverse VQC architectures with varied encoding schemes across multiple medical datasets. We notably show that the choice of ansatz significantly impacts the performance of the algorithm, and start to examine some theoretical clues to elucidate why this is the case. 



\subsection{Performance analysis with theoretical clues}
Different metrics have been proposed in literature to quantify certain properties of quantum circuits, such as the expressibility, the entanglement capability and the effective dimension (\cite{sim_expressibility_2019, abbas_power_2021}). Work to link these properties to the performance of the VQC has not yet shown conclusive correlation. In ~\cite{dragan_quantum_2022}, the authors study the influence of the chosen ansatz on the prediction quality and speed of convergence of a hybrid quantum reinforcement learning algorithm. We studied in \cite{monnet_pooling_2023} the link between various quantum pooling strategies to reduce the dimension of the data and the effective dimension of the circuit, again without finding a correlation between the two. This still raises the question as to whether these metrics might be relevant for choosing a good embedding strategy.

Alternatively, it is possible to look at the problem from a different angle: previous work in (\cite{schuld_effect_2021, vidal_input_2020}) shows that QML models that use the variational approach generate Fourier-type sums. From a signal processing point of view, this means that the outputs of quantum circuits can be decomposed into their constituting sine and cosine waves, each with a weighted contribution to the overall output. In~\cite{casas_multidimensional_2023}, the authors analyze this formalism and its implications for different types of circuit architectures commonly used in QML for both one-dimensional and multi-dimensional inputs, while using a varying number of data encoding and ansatz layers. The authors of~\cite{aikaterini_effect_2022} compare the Fourier coefficient distributions for a 2-qubit circuit and a similar one with an additional ancillary qubit, revealing distinct distributions. This suggests that they learn different function classes despite only varying in qubit count. They also observe that measurements on different qubits in the 2-qubit circuit yield varying coefficient distributions.

To the best of our knowledge, our study represents the first attempt to analyze the performance of a QCCNN using diverse encoding strategies, considering both the quantum metrics mentioned earlier and Fourier decomposition.

\section{Background}
\label{background}
In this section, we provide the relevant information for the construction of the QCCNNs, specifically addressing the encodings employed and the data re-uploading strategy. Following that, we define the quantum metrics used in this work, and describe the Fourier representation of quantum circuits.


\subsection{QCCNNs and encodings within this work}

We build on the study conducted in \cite{matic_quantum-classical_2022} and construct a QCCNN by replacing the first classical convolutional layer of a CNN by a VQC, while keeping the remainder of the architecture unaltered. All qubits are then measured and their outcomes are mapped onto separate output channels. A classical fully-connected layer performs the final classification using the sigmoid activation function.

\begin{figure}
\centering
\begin{quantikz}[scale=0.3]
 & \lstick{$\ket{0}$} &&\gate[4]{U(f * x_{i})}\gategroup[4,steps=1,style={dashed,rounded
corners,fill=blue!20, inner
xsep=2pt},background,label style={label
position=above,anchor=north,yshift=+0.4cm}]{{\sc $S(\textbf{x})$}}&\gate{R_X(\theta_{0})}\gategroup[4,steps=5,style={dashed,rounded
corners, inner xsep=2pt},background,label style={label position=above,anchor=north,yshift=+0.4cm}]{{\sc $W(\boldsymbol{\theta})$}} & \ctrl{1} &\qw&\qw& \targ{1} & \meter{} 
& & 
\\ &\lstick{$\ket{0}$} &\qw&\qw& \gate{R_X(\theta_{1})} &  \targ{2} & \ctrl{1} &\qw&\qw& \meter{}
& &
\\ &\lstick{$\ket{0}$} &\qw&\qw& \gate{R_X(\theta_{2})} &\qw& \targ{2} & \ctrl{1} &\qw & \meter{}
& &
\\ &\lstick{$\ket{0}$} &\qw&\qw& \gate{R_X(\theta_{3})} &\qw&\qw& \targ{2} & \ctrl{-3} & \meter{}
\end{quantikz}
    \vspace{0.5cm}
    \caption{The circuit structure used in this study.}
    \label{fig:VQC}
\end{figure}
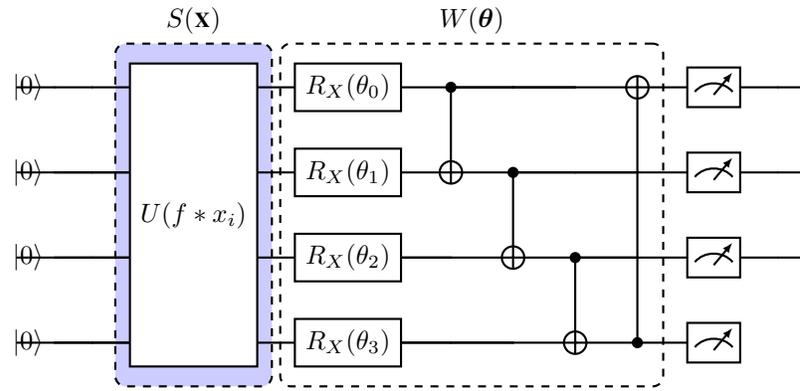

Fig.~\ref{fig:VQC} shows the general structure of this circuit. The VQCs in this study are designed as follows: we first define an encoding strategy $S(x)$, then we select an ansatz $W(\theta)$ to remain unchanged for every encoding tested, as we are interested in observing the effect of the data embedding. We choose a basic entangling layer as it combines simplicity with high performance (\cite{matic_quantum-classical_2022, monnet_pooling_2023}). In this ansatz, each qubit $i$ is rotated by a trainable angle $\theta_{i}$ around the X-axis before applying a sequence of CNOT entangling gates on adjacent qubit pairs. Note that for a 2x2 filter size in the convolutional layer, the number of qubits required to construct such a filter may vary depending on the choice of encoding, e.g. 4 qubits for the angle and higher-order encoding, versus 2 for the amplitude encoding. For all encoding strategies used in this study, a rescaling factor $f$ is multiplied with the input data.

The following encoding techniques are explored:

\begin{enumerate}
    \item \textbf{Angle Encoding}: This embedding method is popular particularly due to its simple implementation and typically satisfactory performance \cite{matic_quantum-classical_2022}. For an N-entry classical data vector $\mathbf{x} = (x_1, .. , x_N)$ we can encode the data as
    \begin{equation}
        \ket{\Psi} = \bigotimes_{i=1}^{N} \text{cos}(x_{i}) \ket{0}+ \text{sin}(x_{i}) \ket{1}
    \end{equation}
    and we can write the state preparation unitary as \(\mathcal{U} = \bigotimes_{i=1}^{N}  exp \left(\frac{-i \phi \sigma_j}{2}\right)\)
    with $\sigma_{j}$ the Pauli matrices, generators of the Pauli gates. While the execution of this encoding is relatively straightforward, there is no efficiency benefit in the number of required qubits, with an N-dimensional input requiring an N-qubit system. In this work, we use two possible angle encoding schemes, namely with an $R_{X}$ gate or an $R_{Y}$ gate, corresponding to $\sigma_{X}$ and $\sigma_{Y}$ respectively.
    \\

    \item \textbf{Higher-Order Encoding}: First proposed in~\cite{havlicek_supervised_2019}, this encoding was initially designed for Support Vector Machine (SVM) classifiers where it was shown that a quantum advantage can only be achieved if the encoding feature map is classically hard to simulate. The feature map unitary as used  in~\cite{abbas_power_2021} is given by
    \begin{equation}  \ket{\Psi} = H^{\otimes n} \exp \left( i \sum_{S\subseteq[n]} \phi_S (\vec{x}) \prod_{i \in S} Z_{i} \right).
    \end{equation}
    A Hadamard gate is first applied on each qubit, followed by a first-order angle encoding using $R_{Z}(\phi_1 (x_{i}))$ gates with $\phi_1 (x_{i}) = f * x_{i}$ and a pairwise second-order operation $R_{ZZ}(\phi_2 (x_{i},x_{j}))$ with $\phi_2 (x_{i},x_{j}) = f* x_{i} x_{j}$.
    \\

    \item \textbf{Amplitude Encoding}: Classical data can be represented as amplitudes of a quantum state. For a  classical data vector $x = (x_1, .. , x_N)$, this can be written as
    \begin{equation}
    \ket{\Psi} = \sum_{i=1}^{N} x_{i} \ket{i}
    \end{equation}
    where $\ket{\Psi}$ is the resulting quantum state, $x_{i}$ are the amplitudes containing the classical information such that \(\sum_{i=1}^{N} |x_{i}|^{2}=1\), and $\ket{i}$ are the quantum states for each vector component, with $i$ being the index written in binary. An N-dimensional amplitude encoded classical vector requires $\log_{2}(N)$ qubits. This implies that we only require 2 qubits to encode an input of size 4. This reduction in the required number of qubits is the main advantage of amplitude encoding. 
    To prepare the amplitudes, a quantum state preparation is used. In this work, we use the Möttönen state preparation~\cite{mottonen_transformation_2004} as implemented in pennylane \cite{bergholm_pennylane_2022}. 

\end{enumerate}

\subsection{Data-reuploading}


Unlike classical neural networks, which are able to process the original data multiple times in their hidden layers, quantum neural networks are limited by the no-cloning theorem. To circumvent this limitation,~\cite{perez-salinas_data_2020} propose a single-qubit universal classifier consisting of an encoding and a trainable unit through which the data is repeatedly passed. For a general single-qubit rotation $U(\theta, \phi, \delta)$ which we consider as the processing unit, and an arbitrary data encoding rotation gate $U(x)$, we can define a single layer as $L(x, \mathbf{\theta}, \mathbf{\phi}, \mathbf{\delta})=U(\theta, \phi, \delta) \; U(x)$. This constitutes the building block of the classifier and can be applied repeatedly. The circuit unitary for $N$ layers can be written as $\mathcal{U}(\textbf{x}, \mathbf{\theta}, \mathbf{\phi}, \mathbf{\delta}) = L_{1}(x_{1},\theta_{1},\phi_{1},\delta_{1}) \; .. \; L_{1}(x_{N},\theta_{N},\phi_{N},\delta_{N})$. The principle can also be extended to a multi-qubit classifier which enables entangling operations between qubits, improving the performance and lowering the number of required layers.

\subsection{Quantum Metrics}
\label{sec:metrics}
Multiple metrics to describe the properties of quantum circuits were proposed in literature.
In this work, we investigate three quantum metrics to identify their possible correlations with the choice of an encoding strategy.
\begin{enumerate}
    \item \textbf{Expressibility}~\cite{sim_expressibility_2019}: The expressibility quantifies the ability of a circuit to explore the Hilbert space. For a single qubit, a highly expressible VQC is able to explore the entire Bloch Sphere uniformly. The degree of uniformity is measured by randomly sampling parameters $\theta$ of the VQC and taking the statistical distance measure Kullback–Leibler-divergence relative to the ensemble of Haar random states. If a circuit is highly expressible, this distance is low. This is given by
    \begin{equation}
        Expr=D_{\mathrm{KL}}\left(\hat{P}_{\mathrm{VQC}}(F~; \boldsymbol{\theta})~\| P_{\text{Haar}}(F)\right)
    \end{equation}
    where the fidelities $F$ are obtained by repeating the procedure of uniformly sampling doublets of parameter vectors and computing their corresponding state overlap. $\hat{P}_{\mathrm{VQC}}(F~;~\boldsymbol{\theta})$ is the estimated probability distribution of fidelities obtained by sampling from the VQC and $P_{\text {Haar}}(F)$ is known and obtained analytically by $P_{\text {Haar}}(F) = (N-1)(1-F)^{N-2}$~\cite{zyczkowski_average_2005}.
    \\
    \item \textbf{Entanglement capability}~\cite{sim_expressibility_2019}: This metric is given by the Meyer-Wallach (MW) entanglement measure and quantifies the capacity of a VQC to generate entangled states. It would be equal to zero for a circuit that generates product states (no entanglement) and one in the case of an output consisting only of maximally entangled states. 
    The MW measure is described by
    \begin{equation}
         Q(|\psi\rangle) \equiv \frac{4}{N} \sum_{j=1}^N D\left(\iota_j(0)|\psi\rangle, \iota_j(1)|\psi\rangle\right)
    \end{equation}
    where $N$ is the number of qubits in the system and $\iota_j(b)$ is a linear mapping that acts on the state in the computational basis ($b\in\{0,1\}$) with the $j$-th qubit absent using the Kronecker delta as
    \begin{equation}
        \iota_j(b)\ket{b_1 \ldots b_N}=\delta_{b b_j} \ket{b_1 \ldots b_{j-1}b_{j+1} \ldots b_N}.
    \end{equation}
    
    The generalized distance $D$ is computed as
    \begin{equation}
    D(|u\rangle,|v\rangle)=\frac{1}{2} \sum_{i, j}\left|u_i v_j-u_j v_i\right|^2.
    \end{equation}
    
    The final numerical value used to estimate the entanglement capacity is computed using the following equation
    \begin{equation}
        Ent=\frac{1}{|S|} \sum_{\boldsymbol{\theta}_i \subset S} Q\left(\left|\psi_{\boldsymbol{\theta}_i}\right\rangle\right).
    \end{equation}
    where the average of the MW measure is taken over the set $S$ of sampled circuit parameter vectors $\boldsymbol{\theta_{i}}$.
    \\
    \item \textbf{Effective Dimension} (ED)~\cite{abbas_power_2021}: The ED is designed to evaluate the information capacity of a statistical model. When considering ML models as such, the actual dimension they explore may be a subset of the model space. The metric characterizing this space is given by the Fisher Information Matrix (FIM). A higher effective dimension can be interpreted as corresponding to a model exploring more of the space of all its possible functions. We define a $d$-dimensional real Riemannian parameter space $\Theta$ for our statistical model $\mathcal{M}_{\Theta}$ with parameters $\theta \in \Theta \subset [-1,1]^{d}$ and consider input-output data pairs $(x,y)$ with $x \in \mathcal{X}^{in}$ and $y \in  \mathcal{Y}^{out}$ which are related for a fixed choice of parameter $\theta$ as $p(x,y;\theta)$ = $p(y|x;\theta)p(x)$. One can capture the amount of information acquired from such parameterization using the FIM $F(\theta) \in \mathbb{R}^{d \times d}$, described by
    \begin{equation}
     F(\theta)=\mathbb{E}\left[\frac{\partial}{\partial \theta} \log p(x, y ; \theta) \frac{\partial}{\partial \theta} \log p(x, y ; \theta)^{\top}\right]
    \end{equation}
    Geometrically, the FIM captures the curvature of the log-likelihood function and it is a gauge for the sensitivity of the model to movements in the parameter space. A high FIM reflects a steeper curvature and thus greater information content. 
    With the help of the FIM as a metric for the parameter space $\Theta$, we can now define the effective dimension as a bounded capacity measure for a number $n>1$ of data samples as
\end{enumerate}
\begin{equation}
    d_{eff}\left(\mathcal{M}_{\Theta}\right):=\frac{2 \log \left(\frac{1}{V_{\Theta}} \int_{\ominus} \sqrt{\operatorname{det}\left(\mathrm{id}_d+\frac{\gamma n}{2 \pi \log n} \hat{F}(\theta)\right)} \mathrm{d} \theta\right)}{\log \left(\frac{\gamma n}{2 \pi \log n}\right)}
\end{equation}
\begin{quote}
In order to obtain a quantity that can be used to compare the different models, the effective dimension is then divided by the number of parameters and is referred to as normalized effective dimension (NED).
\end{quote}

\subsection{Quantum circuits and Fourier series}
One way to understand QML models constructed with VQCs is through a Fourier series representation \cite{schuld_effect_2021}. This is especially relevant for models that use data-reuploading schemes with encoding gates of the form $S(x) = e^{ixH}$, where $x \in \mathbb{R}^M$ is the input and $H$ is a general Hamiltonian for an arbitrary dimension $M \in \mathbb{N}$ of the input data. The function describing the model can thus be written as
\begin{equation}
    f_{\boldsymbol{\theta}}(\mathbf{x}) = \sum_{\mathbf{w}\in \Omega} c_{\mathbf{w}}(\boldsymbol{\theta}) e^{i\mathbf{wx}},
\end{equation}
such that \textbf{x} is a M-dimensional input vector, \textbf{w} is a M-dimensional frequency vector in the $\Omega \subset \mathbb{R}^{N}$ frequency spectrum determined by the eigenvalues of the data encoding Hamiltonians, \textbf{w}\textbf{x} is a scalar product and the coefficients $c_{\mathbf{w}} \in \mathbb{C}$ are determined by the remainder of the circuit and satisfy the condition $c_{\mathbf{w}} = c_{-\mathbf{w}}^{*}$.

In order to extend the frequency spectrum of a quantum circuit and therefore to increase the degree of the Fourier series it generates, one may increase the number of qubits in the circuit, or the number of reuploading layers. 
In the ideal case, a chosen VQC architecture should be able to generate a rich frequency spectrum and adapt the coefficients of the series in such a way that it can approximate the target function well. However, not all architectures are able to contain enough adaptable degrees of freedom to appropriately adjust the coefficients of the series they generate. A detailed study of how the choice of the model architecture affects the number of parameters can be found in~\cite{casas_multidimensional_2023}. In this work, we are solely focused on what the authors refer to as a Parallel Ansatz (PA), where each data feature is encoded in
a different qubit with a single-qubit gate.
Following the notations used in~\cite{casas_multidimensional_2023}, the number of parameters that the PA contains is given by
\begin{equation}
    N_{p}^{PA} = (d^{2M}-1)(L+1)
    \label{eq:params}
\end{equation}
where $d$ is the local dimension given by $d = 2^n$ for a qubit system in which $n$ is the dimensionality of the local space and depends on the encoding, $M$ is the number of qubits, and $L$ is the number of layers.
The degrees of freedom for a general Fourier series of degree $D$ is given by 
\begin{equation}
    \nu = (2D+1)^{M}
    \label{eq:degrees_of_freedom}
\end{equation}
with $D=(d-1)L$. The condition $N_{p}^{PA} \ge \nu$ is to be met for the circuit to adapt all coefficients in the series.

\section{Results and discussion}
\label{results}
We proceed to present the performances of our developed QCCNNs on two medical datasets, considering different encodings and varied scalings of the inputs. Subsequently, we analyze these performances considering three quantum metrics typically employed in literature, and using the Fourier formalism, which describes the function class that quantum circuits give rise to. All experiments in this work were performed in simulation using the PennyLane software library \cite{bergholm_pennylane_2022} in conjunction with PyTorch \cite{paszke_pytorch_2019} in a noiseless environment.

\subsection{Datasets}

In this study, we work with relatively small datasets. This requirement comes both from the application side, as medical imaging datasets usually are in the order of 100 to 1000 images, and from the method side, since quantum variants of CNNs were shown to generalize well even when little training data is available. We select two datasets from the the MedMNIST benchmark (\cite{yang_medmnist_2021,yang_medmnist_2023}).

The first dataset is BreastMNIST, which contains 546 training and 78 validation images of 2D ultrasounds of the breast in a 28x28 pixels resolution. The images display breast tissues that are either normal, malignant or benign. The task is to perform binary classification, where the benign and normal lesions are combined into one class. 

Our second dataset is OrganAMNIST, which consists of cropped and resized images to a 28x28 resolution based on 3D abdomen CT scans from the Liver Tumor Segmentation Benchmark (LiTS) \cite{bilic_liver_2023}. A subset of 1,000 training and 600 validation images is taken for this study. The task is to perform multi-class classification of 11 organs.

\subsection{Performances of the QCCNNs}
\label{sec:performance}

\subsubsection{Performance with varying input scalings}
\label{scalings}
The scaling of the classical input data before encoding is not a trivial choice and impacts both the function expressed by the circuit \cite{schuld_quantum_2019} and the circuit's coverage of the Hilbert space. In this work, the pixel values of the images are normalized between -1 and 1 before a scaling factor $f$ is applied. Due to the excessively long training times, the study of the performance of the QCCNN across different input scaling factors is only conducted on our bigger dataset -- OrganAMNIST -- utilizing one encoding method -- the higher-order encoding -- due to its superior performance in prior research \cite{matic_quantum-classical_2022}.

\begin{figure}[h!]
\centering
\includegraphics[width=0.4\textwidth]{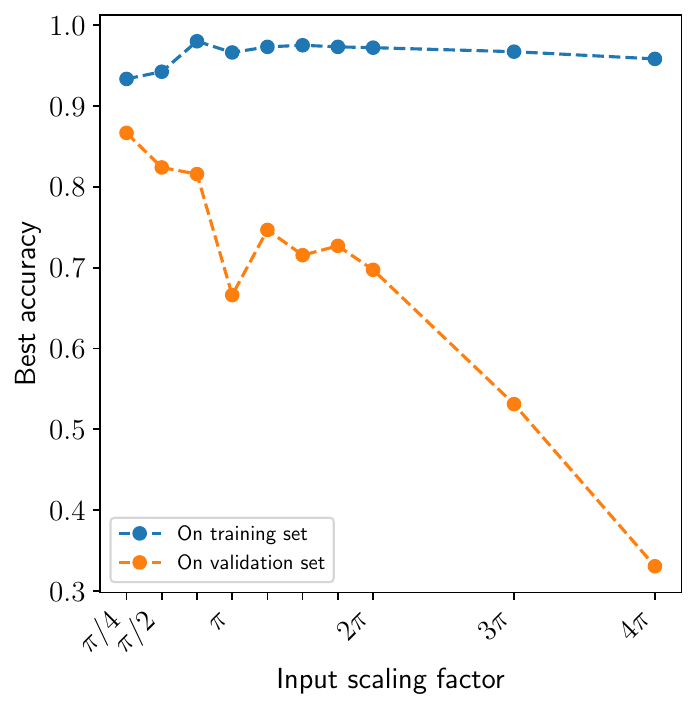}
    \caption{Comparison of the performance of a hybrid QCCNN for different scaling factors on OrganAMNIST using the higher-order encoding.}
    \label{fig:input_scaling_factor}
\end{figure}

Fig. \ref{fig:input_scaling_factor} shows the evolution of the best training and best validation accuracies achieved within 20 training epochs for scaling factors in the encoder ranging from $f = \pi/4$ to $f = 4 \pi$. Note that at $\pi$ and given the normalization, the rotation gates already cover the full circle. However, for experimental purposes, we extend to $4 \pi$. This figure clearly shows that the validation accuracy drops dramatically as the scaling factor grows. While this could be expected for values above $\pi$, this result is surprising, as one would expect the performance to be best when the coverage of the Hilbert space of the circuit is high. Further interpretation on this phenomenon will be provided in section~\ref{sec:quantum_metrics}. Interestingly, a good training accuracy can nevertheless be obtained with all tested scaling factors. In the rest of this paper, a scaling factor of $\pi/4$ was selected for all encodings to maximize the obtained accuracies.

\subsubsection{Performance with varying encodings}
\label{performance}
\begin{figure*}
\centering
\begin{subfigure}{0.4\textwidth}
    \centering
    \includegraphics[width=\textwidth]{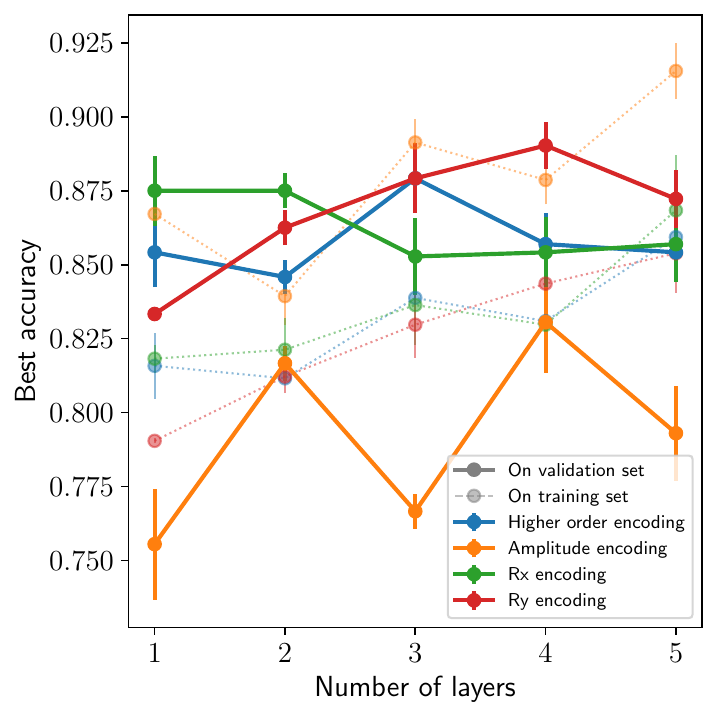}
    \caption{On the BreastMNIST dataset}
\end{subfigure}
 \qquad
\begin{subfigure}{0.4\textwidth}
    \centering
    \includegraphics[width=\textwidth]{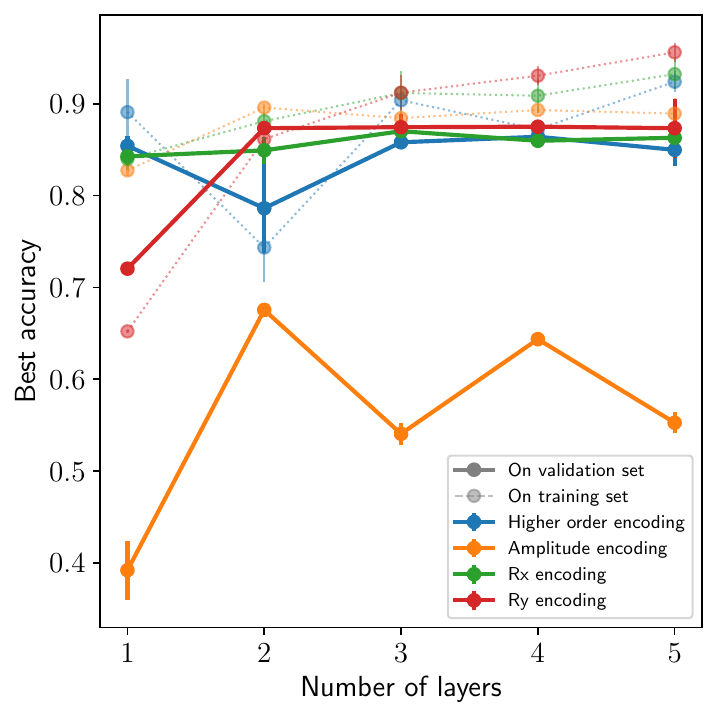}
    \caption{On the OrganAMNIST dataset}
\end{subfigure}
\qquad
\caption{Comparison of the performance of 4 encoding methods for up to 5 layers for hybrid QCCNNs in terms of best training and validation accuracies.}
\label{fig:best_accuracy}
\end{figure*}

We proceed to compare the performance of the developed QCCNNs for the chosen encoding methods using data-reuploading.
The performances are analyzed in terms of best reached training and validation accuracies achieved within 20 training epochs. For every QCCNN architecture, three training sessions were conducted with different weight initialization. The reported metrics represent the averages across these training runs, and the error bands reflect the variance observed.
Fig. \ref{fig:best_accuracy} compares the accuracies achieved by the models with the $R_{X}$, $R_{Y}$, higher-order, and amplitude encodings for 1 up to 5 reuploading layers on BreastMNIST and OrganAMNIST. 

With one encoding layer, the higher-order and $R_{X}$ encoders achieve the best validation accuracy on both datasets, even when taking into account the error bands that are notably larger on the BreastMNIST dataset, due to the smaller size of the dataset. The $R_{Y}$ encoding performs a little bit worse, while the amplitude encoding performs significantly worse in validation accuracy, despite achieving a training accuracy on-par with the other models. This gap in performance of the architecture with amplitude encoding is likely attributed to its utilization of a smaller number of qubits (2 instead of 4 as seen in other approaches).

With two layers, the higher-order encoder exhibits a slight decline compared to other methods, while the $R_{X}$ encoder remains stagnant, and the $R_{Y}$ encoder achieves validation accuracies comparable to (on BreastMNIST) or even surpasses the two previous methods (on OrganAMNIST). The amplitude encoder benefits from the additional reuploading layer but the performance remains less than ideal.

When adding three and more reuploading layers, the gain and loss in accuracy due to the additional layers seems to saturate as the accuracies remain more or less constant on OrganAMNIST for all architectures but the amplitude encoding. In particular, they reach a similar final accuracy around 87\%. On BreastMNIST, in spite of the relatively large error bands, we observe a similar behavior of saturation of the performance. On the other hand, the architecture with amplitude encoding alternatively varies as the number of reuploading layers increases, while overall improving but remaining at a low value compared to the other encoders.

All in all, the highest mean validation accuracy in all considered models is achieved on both datasets with the $R_{Y}$ encoding, at 4 and 2 layers for the BreastMNIST and OrganAMNIST, respectively.

\subsection{Quantum metrics and their correlation to performance}
\label{sec:quantum_metrics}
\begin{figure*}
\centering
\begin{subfigure}{0.45\textwidth}
    \centering
    \includegraphics[width=\textwidth]{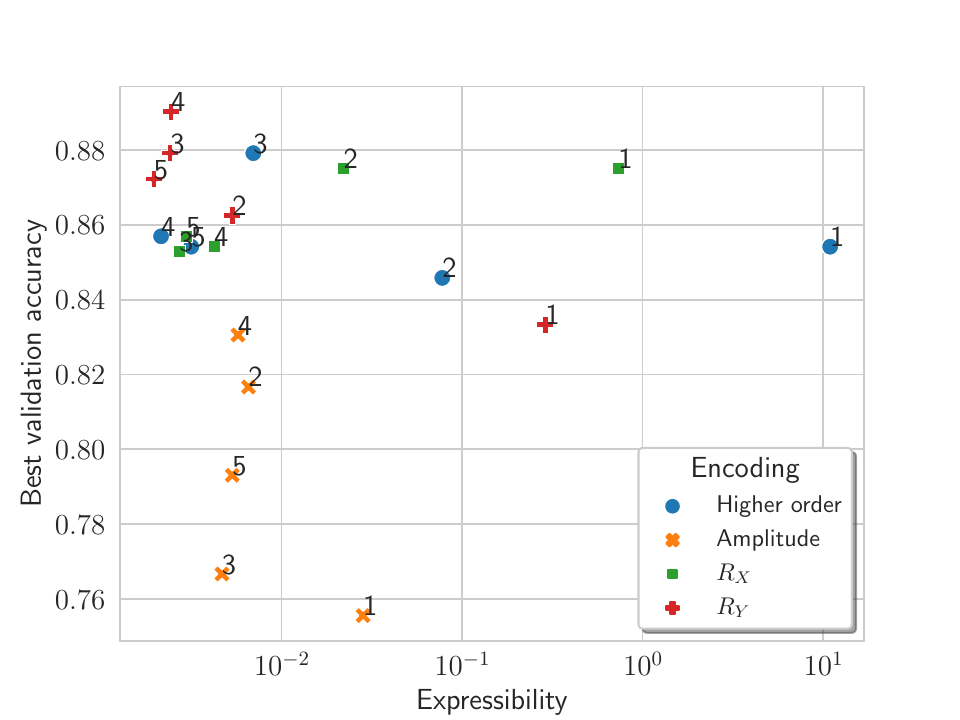}
    \caption{Expressibility on the BreastMNIST dataset}
\end{subfigure}
 \qquad
\begin{subfigure}{0.45\textwidth}
    \centering
    \includegraphics[width=\textwidth]{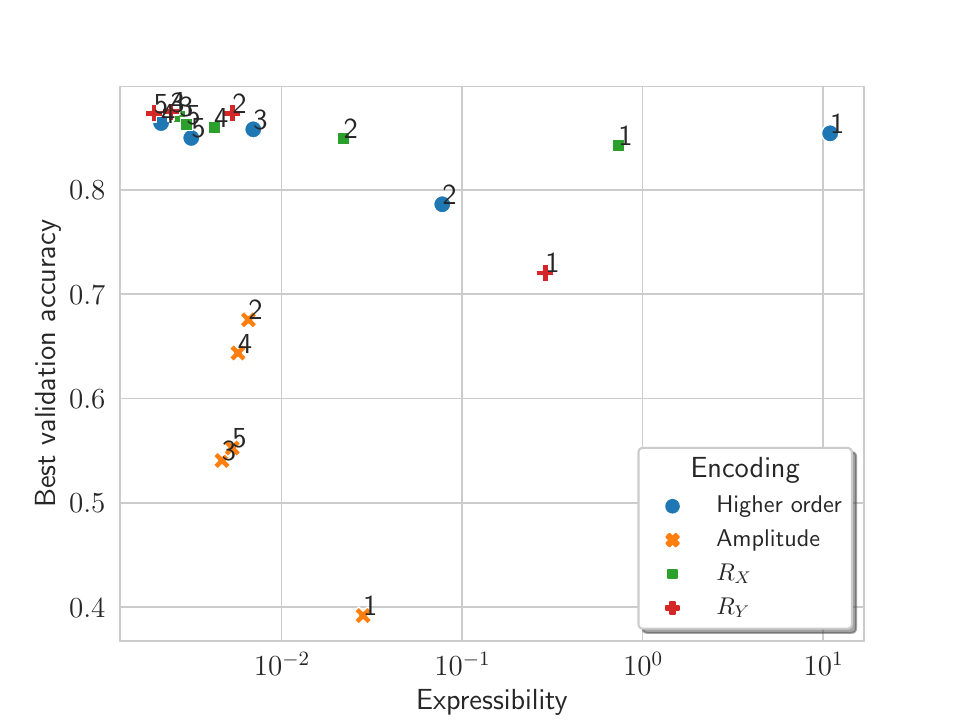}
    \caption{Expressibility on the OrganAMNIST dataset}
\end{subfigure}
\qquad

\begin{subfigure}{0.45\textwidth}
    \centering
    \includegraphics[width=\textwidth]{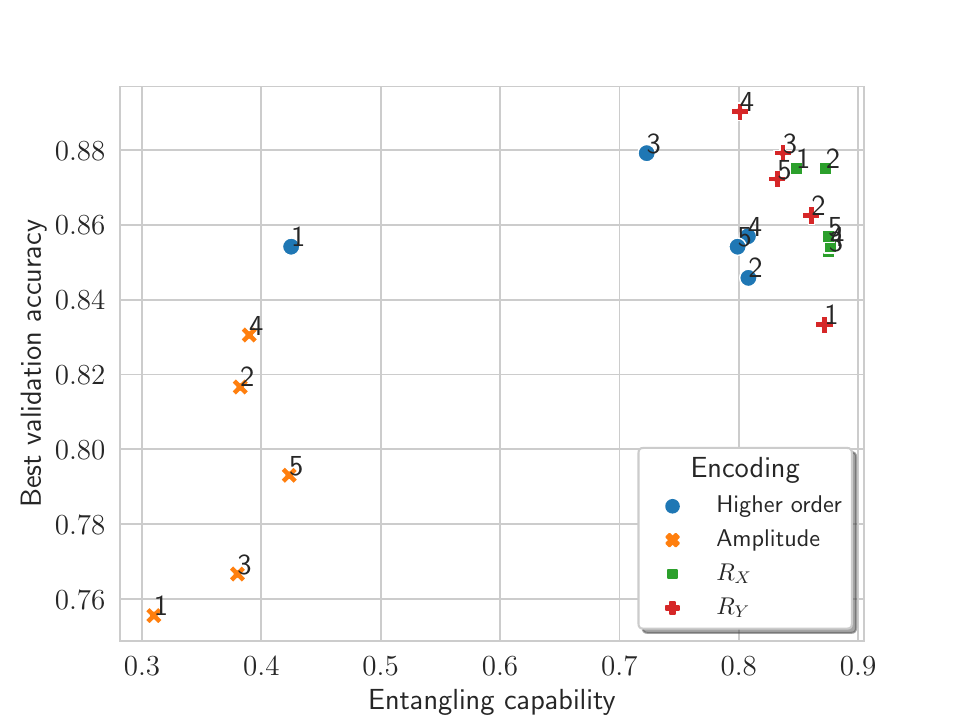}
    \caption{Entangling capability on the BreastMNIST dataset}
\end{subfigure}
 \qquad
\begin{subfigure}{0.45\textwidth}
    \centering
    \includegraphics[width=\textwidth]{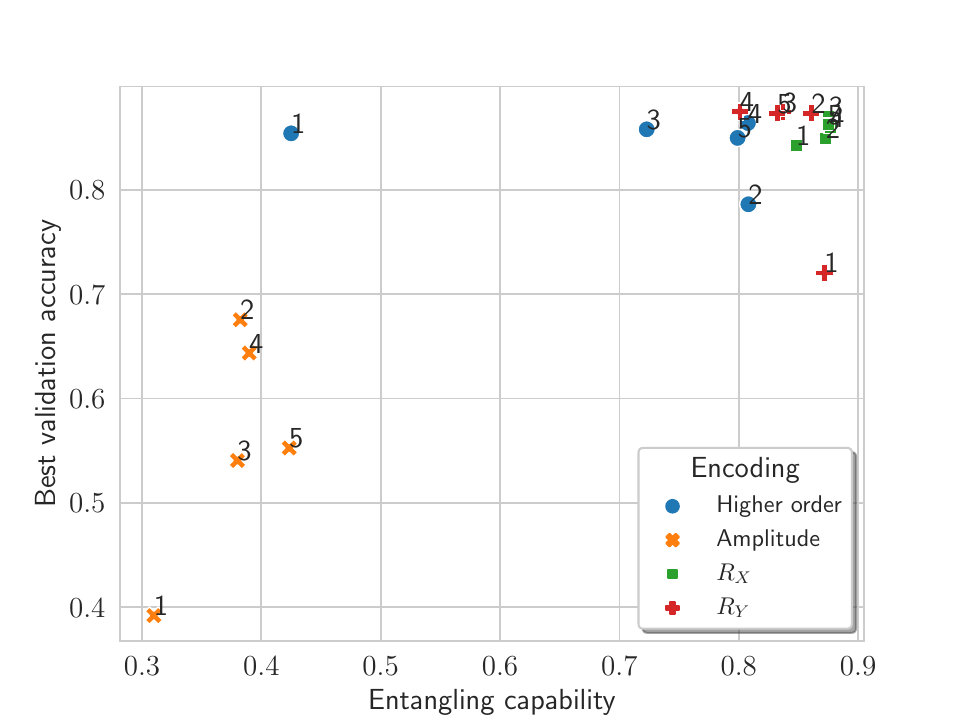}
    \caption{Entangling capability on the OrganAMNIST dataset}
\end{subfigure}
\qquad

\begin{subfigure}{0.45\textwidth}
    \centering
    \includegraphics[width=\textwidth]{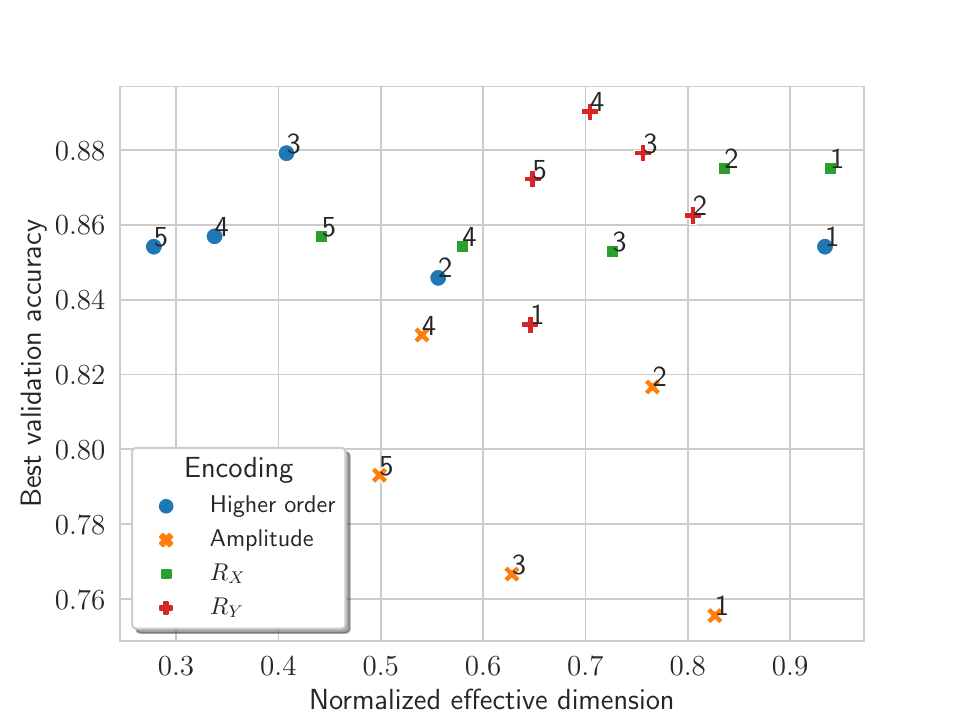}
    \caption{NED on the BreastMNIST dataset}
\end{subfigure}
 \qquad
\begin{subfigure}{0.45\textwidth}
    \centering
    \includegraphics[width=\textwidth]{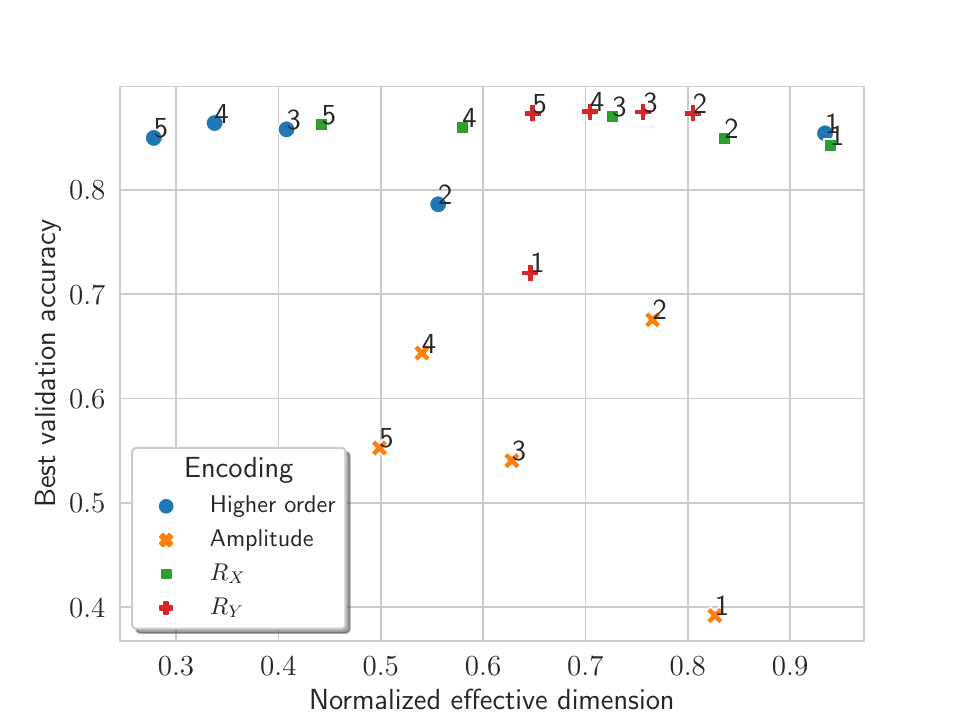}
    \caption{NED on the OrganAMNIST dataset}
\end{subfigure}
\qquad
\caption{Correlation plots between the performance of 4 encoding methods for up to 5 layers and the expressibility, the entangling capability and the normalized effective dimension. The numerical labels indicate the number of layers.}
\label{fig:metrics}
\end{figure*}

We now examine the various metrics outlined in section \ref{sec:metrics} and explore potential correlations with the performance presented in \ref{performance}. The highest validation accuracy attained for all models is plotted against the analyzed quantum metrics in Fig. \ref{fig:metrics}. We first note that these correlation plots look similar for both datasets.

Starting with the expressibility, we see that most circuits exhibit low values, making them highly expressible, with the exception of the $R_{X}$, $R_{Y}$ and higher-order encodings with one layer, which have comparatively higher values. Increasing the number of data-reuploading layers results in an increase in the expressiblity, but is not necessarily accompanied with an improved validation accuracy. Among the more expressible circuits, we encounter instances with good accuracy such as the four-layer $R_{Y}$ encoding, while others, such as amplitude encoding with 4 layers, demonstrate comparatively poorer performance. This tends to indicate that there is no clear correlation between the best achieved accuracy and the expressibility.

It should additionally be noted that in the experiment outlined in section \ref{scalings}, where a fixed encoding strategy, the higher-order encoding, was employed, the accuracy notably decreased with an increase in the scaling factor. However, a higher scaling factor corresponds to a more expressible circuit as it allows it to encompass a larger portion of the Hilbert space. This observation suggests a negative correlation between expressibility and performance achieved. Such an effect may stem from training challenges associated with high expressibility, as discussed in \cite{holmes_connecting_2022}. The study highlights that low-expressibility ansätze (characterized by a high expressibility value) can result in both small and large cost gradients, whereas high-expressibility ansätze (low expressibility value) tend to exhibit predominantly flat cost landscapes, making them generally challenging to train.

Next, we examine the entanglement capacity of the various encodings. One first notes that the amplitude encoding displays clearly lower entangling capabilities than other setups. This is expected as only 2 qubits are used in this case. For the other architectures, one observes a cluster around high values of entanglement capacity, with the exception of the higher-order encoding with one layer. No correlation is therefore observable in the plot between the best validation accuracy and the entanglement capability. For instance, both the four-layer $R_{Y}$ and the two-layer higher-order encoding settings exhibit similar metric values but show significantly divergent performances across both datasets. 


Lastly, the normalized effective dimension appears to follow a downward trend as the number of layers increases, with the exception of the $R_{Y}$ encoding, where an increase in NED is observed from layer 1 to layer 2, coinciding with a significant increase in validation accuracy. Although no clear correlation is observed between the NED and the VQCs' performances, these results seem to indicate a saturation effect as the number of layers increases, given four qubits and the chosen application datasets, as suggested in \cite{casas_multidimensional_2023}. 


\subsection{Fourier Coefficients and their correlation to performance}
In the following subsection, the Fourier coefficients generated by the angle encodings and the higher-order encoding are examined, and potential relations to the performance are discussed. The amplitude encoding is not addressed here due to the state normalization requirement. To address visualization challenges associated with higher-dimensional cases, this analysis is done in a univariate scenario, where the same input is passed to each qubit in the quantum circuit. 
The output is classically decomposed into its constituent frequencies using the discrete Fourier decomposition. The coefficients corresponding to the various frequencies can then be displayed on the complex plane-where the x-axis represents the real part and the y-axis is the imaginary component-and analyzed.
This study is a simplification of what really happens in the quantum convolution, where at each iteration of the convolution process, one pixel value is mapped onto one qubit, and where therefore all mapped values differ. This univariate analysis neglects the possibility of having different input pixel values at a given stride of the filter.
Nonetheless, it provides some insights into the expressive power of the different circuits.

\subsubsection{Experimental procedure}

A set of 21 ordered, linearly spaced inputs within $[-1,1]$ are sampled and passed onto the circuit. This replicates the possible input values in the normalized datasets. 100 weights are additionally randomly sampled from $[0, 2 \pi[$ from a uniform distribution. The outputs are then measured and decomposed using the Real Fast Fourier Transform (RFFT) implemented in the Numpy package \cite{harris_array_2020}. The chosen number of coefficients to be considered was set to 10, plus the zero frequency coefficient. 
The distribution of the resulting coefficients is displayed in the complex plane. For each encoding in this analysis, the procedure is performed for up to four data
reuploading layers. The effect of the chosen embedding strategy as well as the increase in the number of layers is studied.



Following the analysis presented in \cite{casas_multidimensional_2023} and by putting equations \ref{eq:params} and \ref{eq:degrees_of_freedom} together, the number of adaptable degrees of freedom is expected to saturate at two circuit layers for the two angle encodings (where the dimensionality of the local system $n = 1$), and at four layers for the higher-order encoding (where $n = 2$ due to the pairwise CNOT operations in the encoding). 

\subsubsection{Fourier coefficients analysis}

\begin{figure*}
\centering
\begin{subfigure}{\textwidth}
    \centering
    \begin{subfigure}{0.47\textwidth}
        \renewcommand\thesubfigure{1.a}
        \centering
        \includegraphics[width=\textwidth]{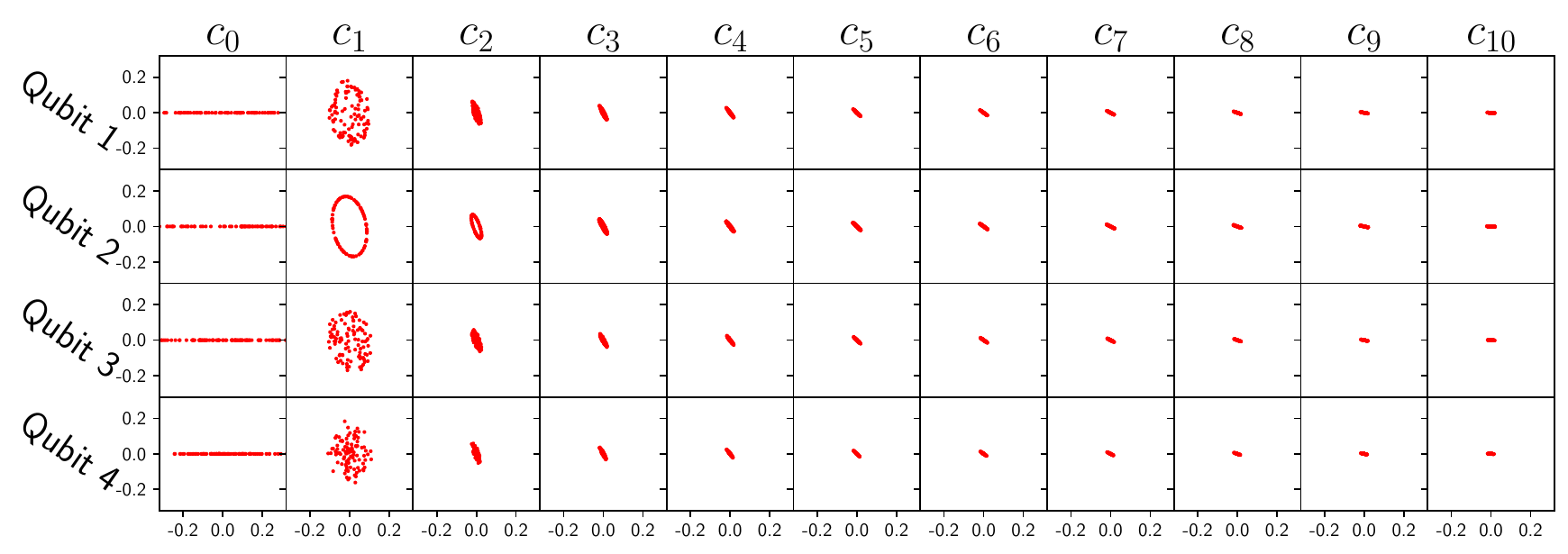}
        \caption{1 layer}
    \end{subfigure}
    \qquad
    \begin{subfigure}{0.47\textwidth}
        \renewcommand\thesubfigure{1.b}
        \centering
        \includegraphics[width=\textwidth]{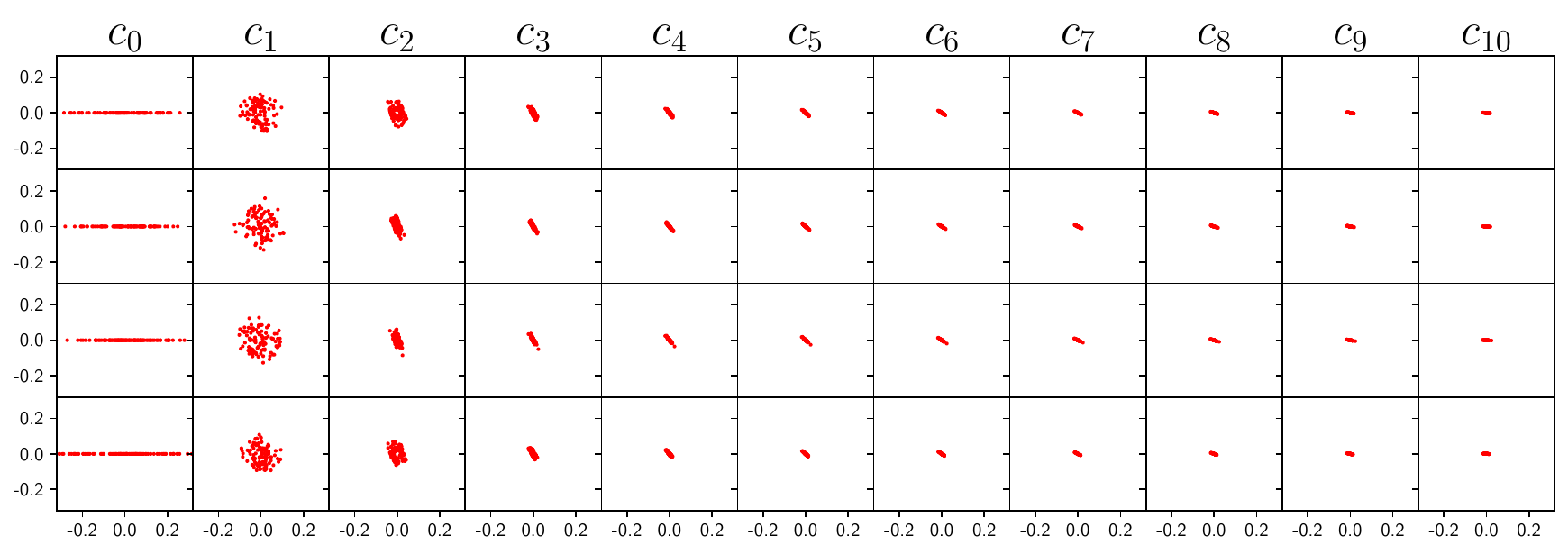}
        \caption{2 layers}
    \end{subfigure}

    \begin{subfigure}{0.47\textwidth}
        \renewcommand\thesubfigure{1.c}
        \centering
        \includegraphics[width=\textwidth]{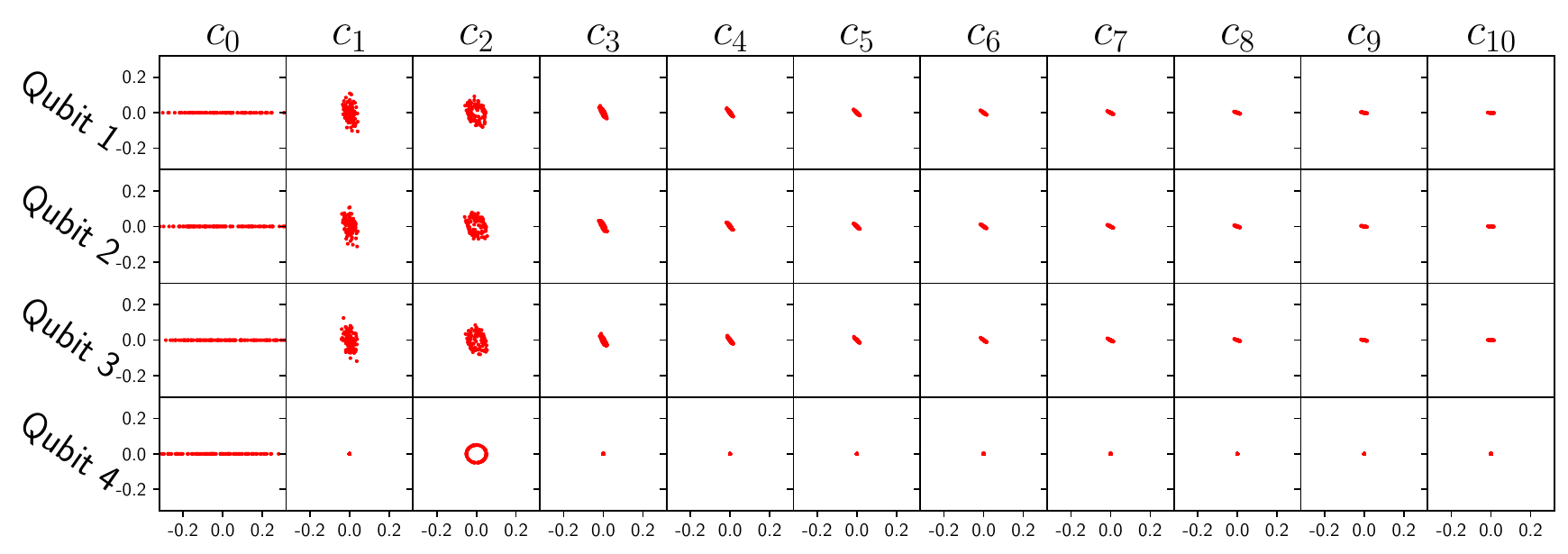}
        \caption{3 layers}
    \end{subfigure}
    \qquad
    \begin{subfigure}{0.47\textwidth}
        \renewcommand\thesubfigure{1.d}
        \centering
        \includegraphics[width=\textwidth]{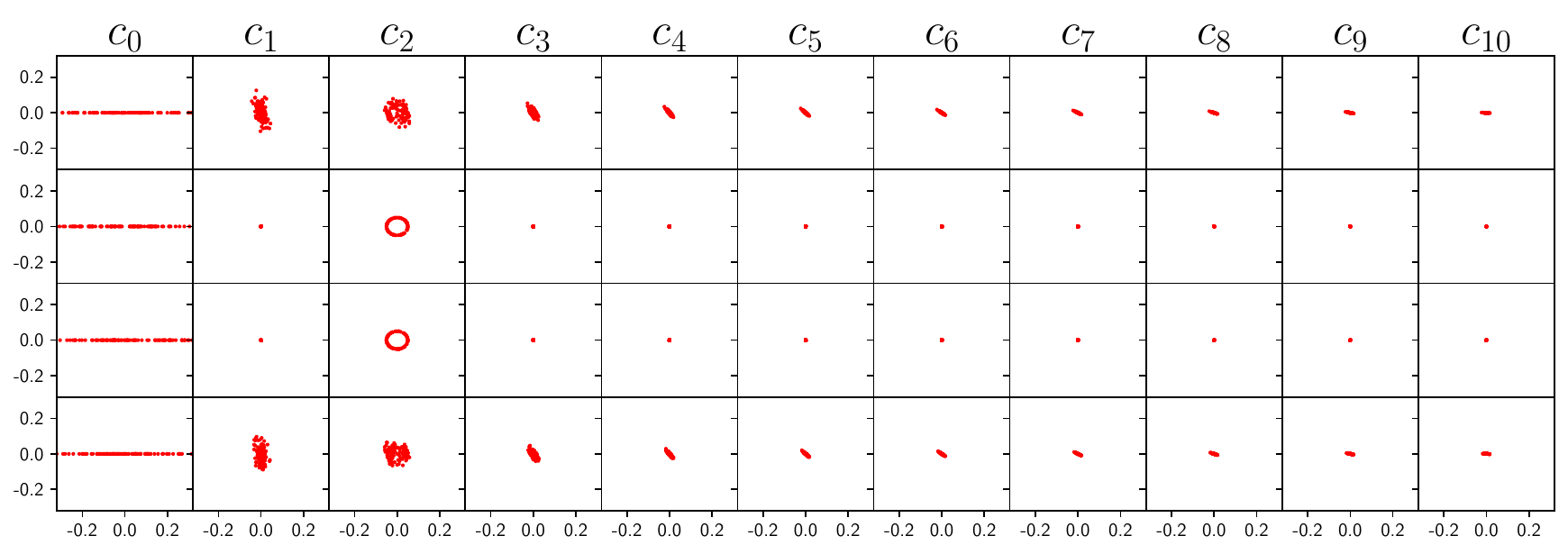}
        \caption{4 layers}
    \end{subfigure}
    \qquad
    \renewcommand\thesubfigure{1}
    \caption{$R_{X}$ encoding}
    \label{fig:uni_$R_{X}$}
\end{subfigure}

\begin{subfigure}{\textwidth}
    \centering
    \begin{subfigure}{0.47\textwidth}
        \renewcommand\thesubfigure{2.a}
        \centering
        \includegraphics[width=\textwidth]{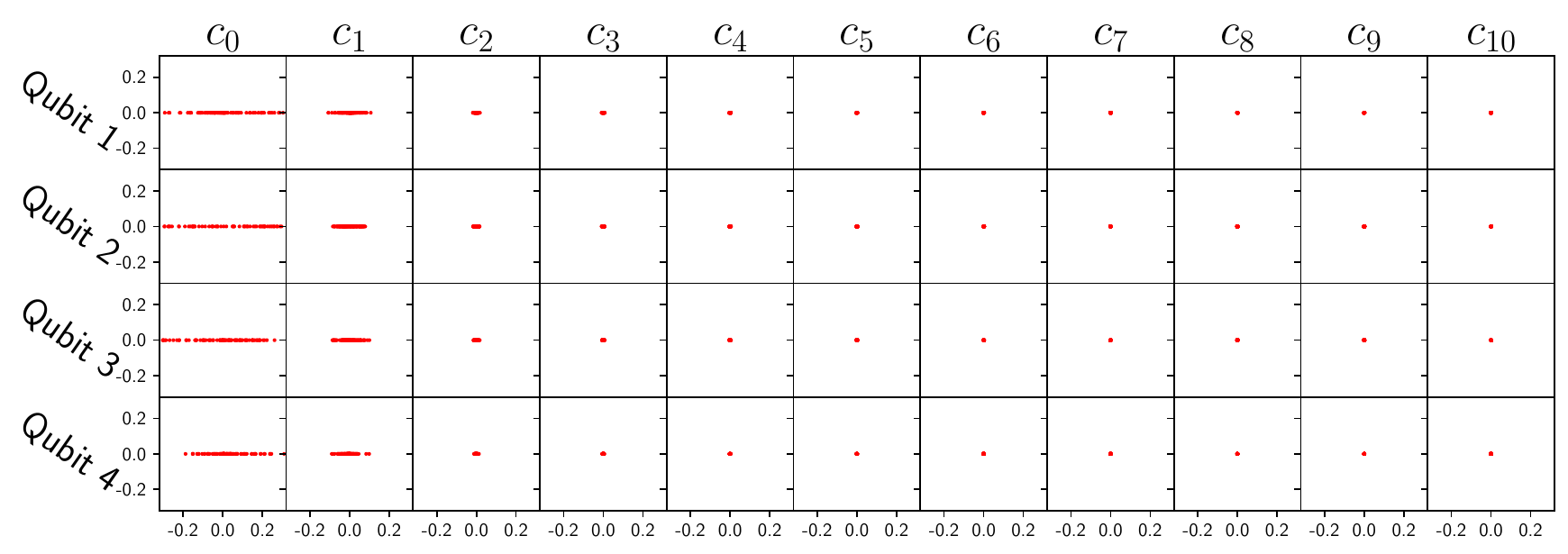}
        \caption{1 layer}
    \end{subfigure}
    \qquad
    \begin{subfigure}{0.47\textwidth}
        \renewcommand\thesubfigure{2.b}
        \centering
        \includegraphics[width=\textwidth]{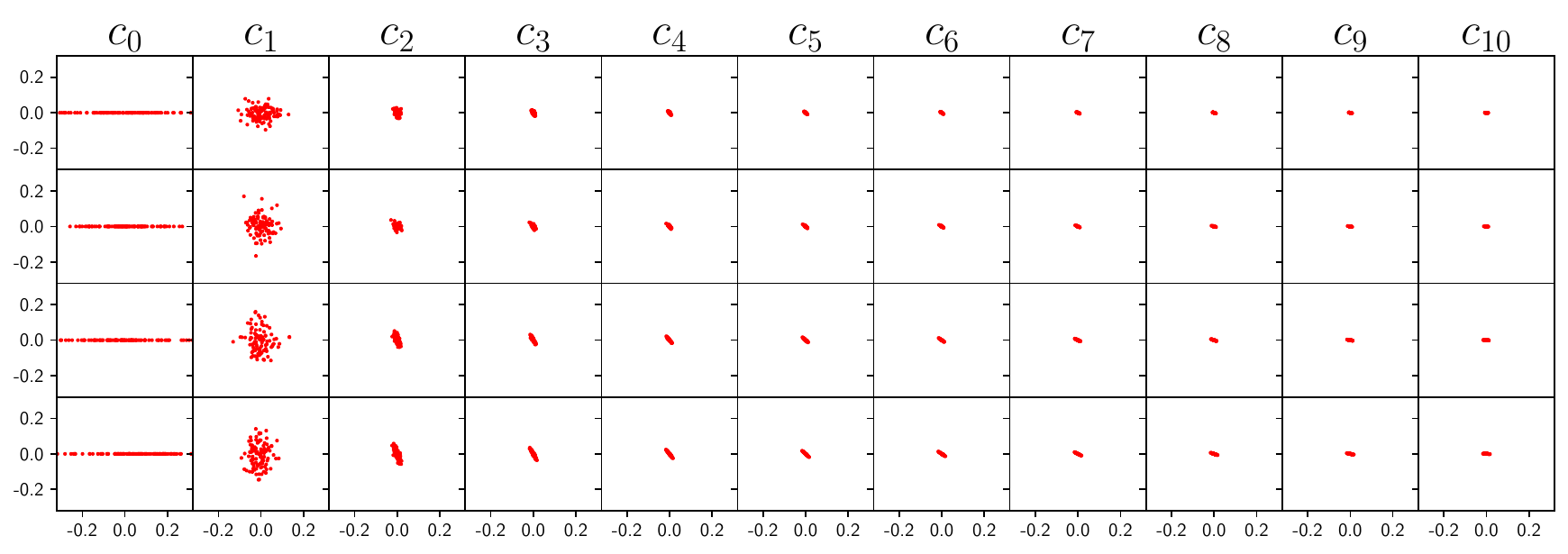}
        \caption{2 layers}
    \end{subfigure}

    \begin{subfigure}{0.47\textwidth}
        \renewcommand\thesubfigure{2.c}
        \centering
        \includegraphics[width=\textwidth]{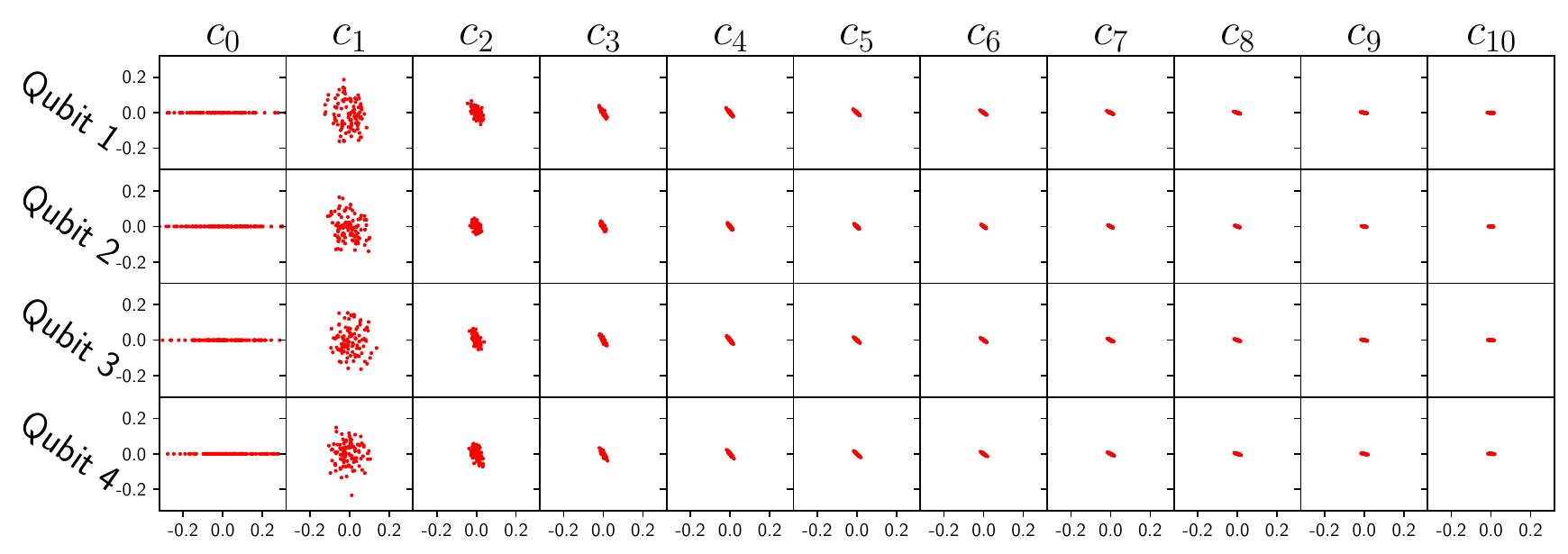}
        \caption{3 layers}
    \end{subfigure}
    \qquad
    \begin{subfigure}{0.47\textwidth}
        \renewcommand\thesubfigure{2.d}
        \centering
        \includegraphics[width=\textwidth]{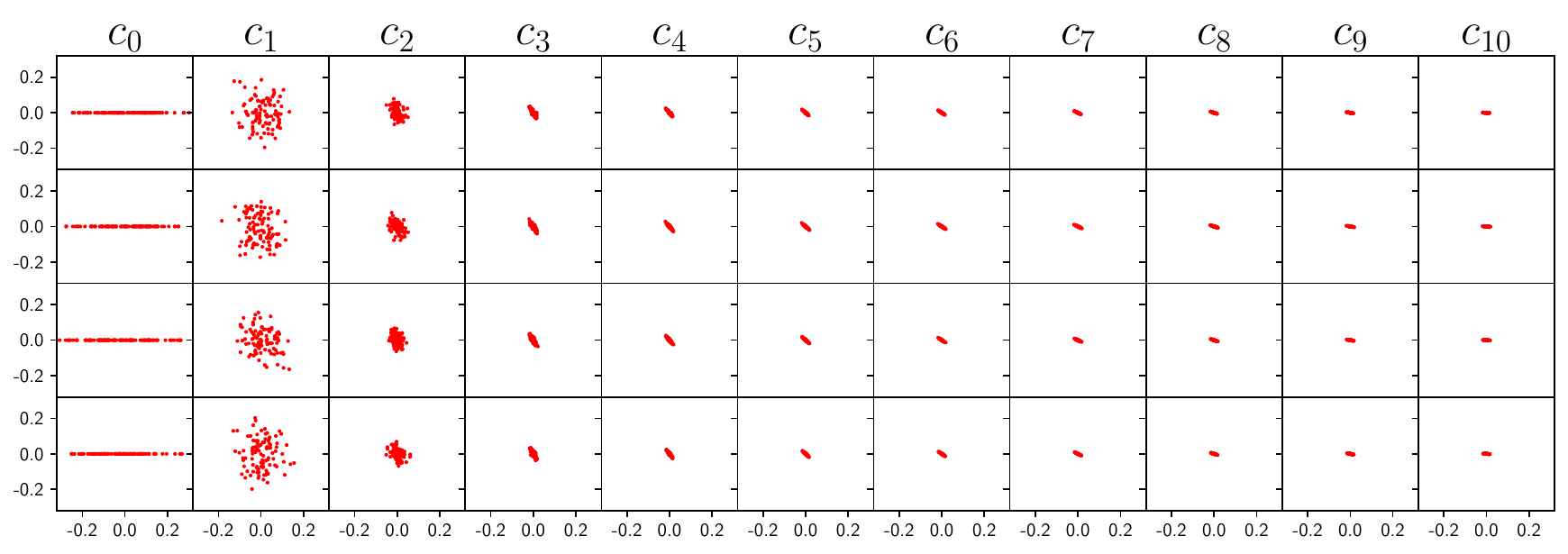}
        \caption{4 layers}
    \end{subfigure}
    \qquad
    \renewcommand\thesubfigure{2}
    \caption{$R_{Y}$ encoding}
    \label{fig:uni_ry}
\end{subfigure}

\begin{subfigure}{\textwidth}
\centering
    \begin{subfigure}{0.47\textwidth}
        \renewcommand\thesubfigure{3.a}
        \centering
        \includegraphics[width=\textwidth]{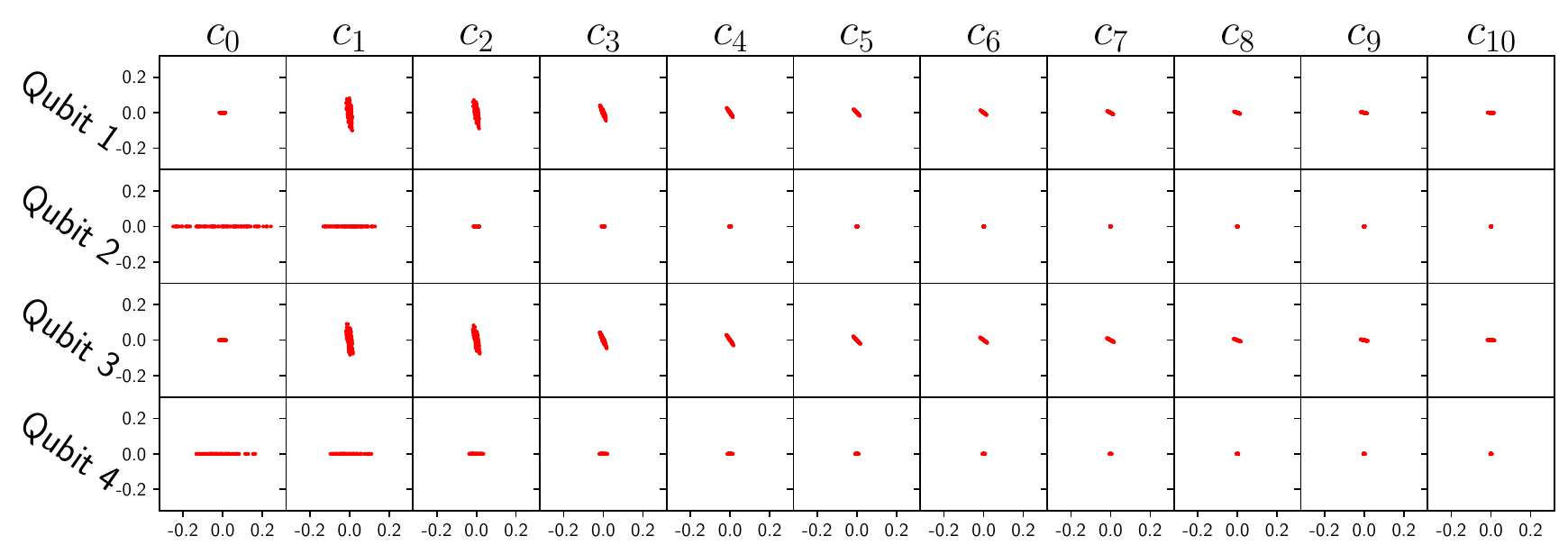}
        \caption{1 layer}
    \end{subfigure}
    \qquad
    \begin{subfigure}{0.47\textwidth}
        \renewcommand\thesubfigure{3.b}
        \centering
        \includegraphics[width=\textwidth]{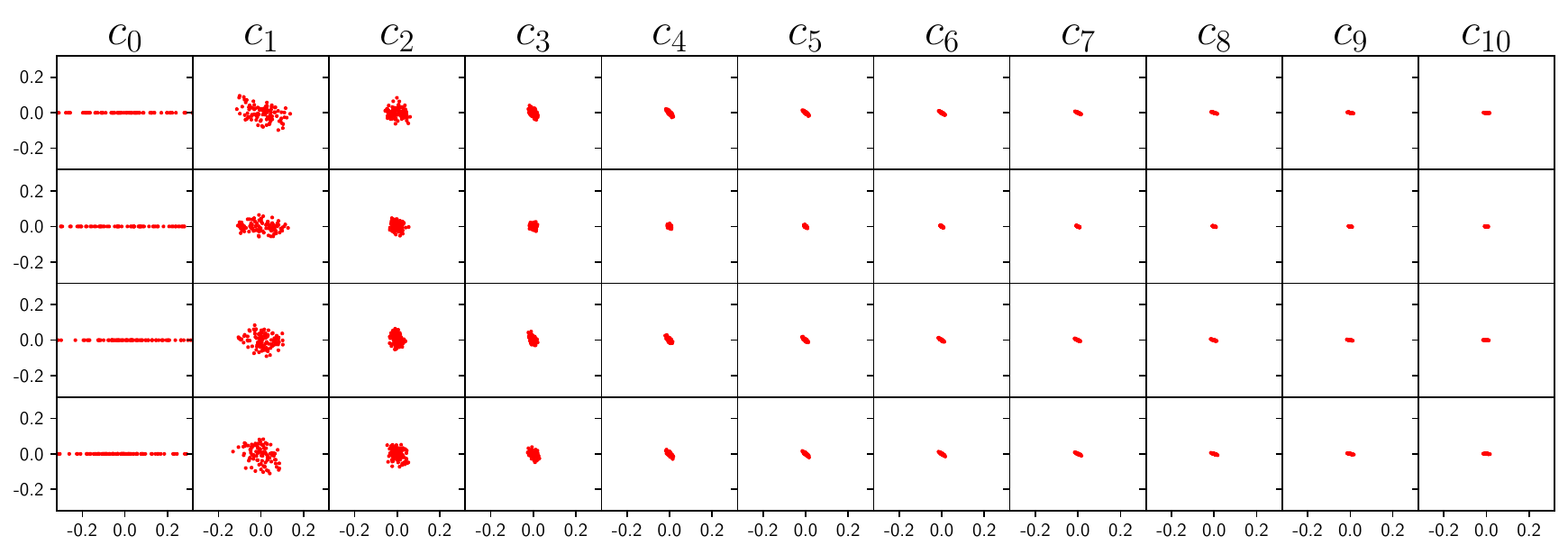}
        \caption{2 layers}
    \end{subfigure}

    \begin{subfigure}{0.47\textwidth}
        \renewcommand\thesubfigure{3.c}
        \centering
        \includegraphics[width=\textwidth]{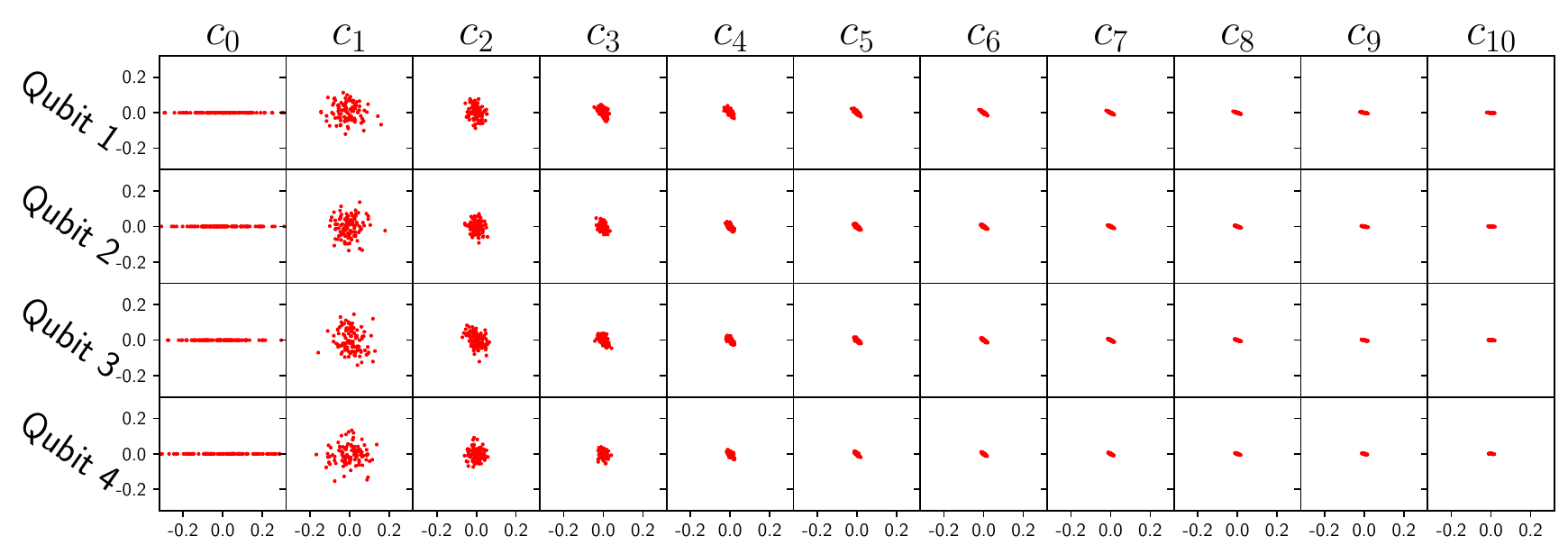}
        \caption{3 layers}
    \end{subfigure}
    \qquad
    \begin{subfigure}{0.47\textwidth}
        \renewcommand\thesubfigure{3.d}
        \centering
        \includegraphics[width=\textwidth]{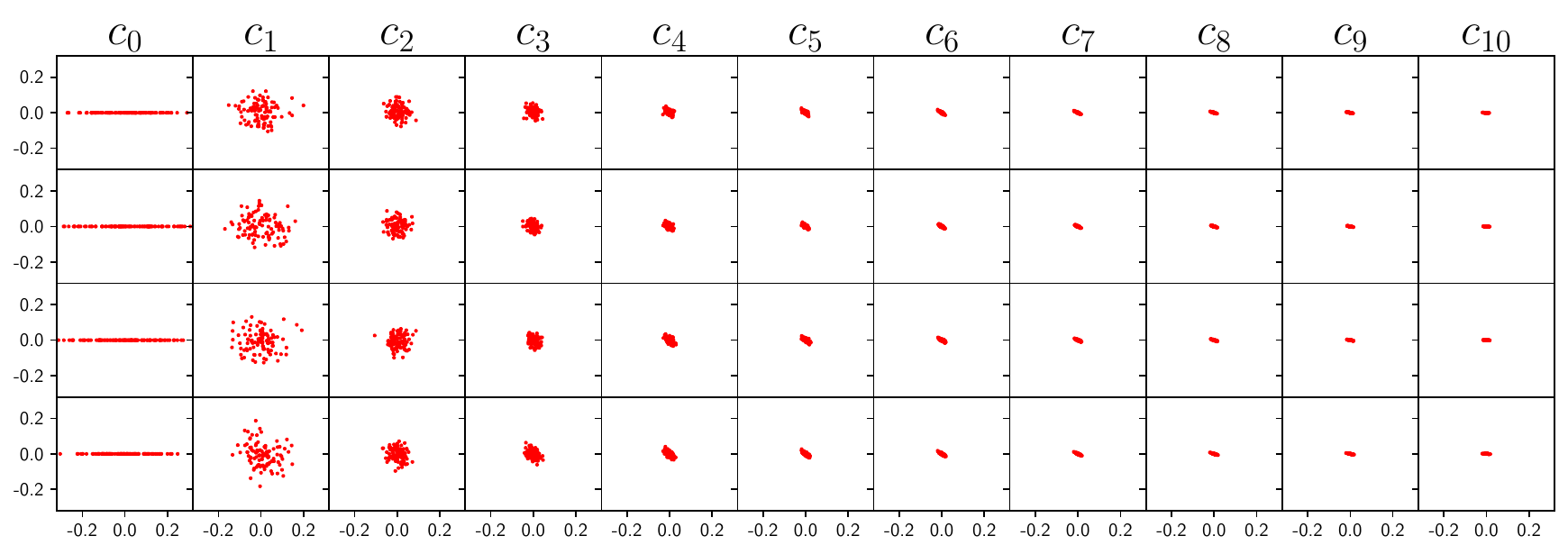}
        \caption{4 layers}
    \end{subfigure}
    \qquad
    \renewcommand\thesubfigure{3}
    \caption{Higher-order encoding}
    \label{fig:uni_higher_order}
\end{subfigure}
\caption{Distributions of the Fourier coefficients of the outputs generated by the various encodings and basic entangling circuit for different reuploading layers.}
\label{fig:fourier}
\end{figure*}

The distributions of the Fourier coefficients for each qubit for the $R_{X}$,  $R_{Y}$ and higher-order encodings are shown in Fig. \ref{fig:fourier}. We particularly pay attention to the number of non-null frequencies present, their rank (from 0 to 10) and the distribution of their coefficients.

At one layer, we first observe that more frequencies are present in the $R_{X}$ and higher-order encodings compared to the $R_{Y}$ encoding, as for $R_{Y}$, the number of non-null Fourier coefficients is limited to the first 4, as opposed to all 10 for the $R_{X}$ encoding and also 10 for the higher-order encoding on qubits 1 and 3. In terms of their coefficient, while a mix of both real and imaginary is accessible with the $R_{X}$ and higher-order encodings (both on the x and y-axis), the  $R_{Y}$ encoding is limited to purely real values (only on the x-axis).
When comparing to the performance of the QCCNNs, we see that on both datasets,  $R_{Y}$ performs worse than $R_{X}$ and the higher-order encodings at 1 layer. This seems to be in line with the interpretation of the expressibility of the Fourier coefficients.

In the $R_{X}$ angle encoding, the addition of one data-reuploading layer results in a more widespread distribution of the first coefficient at qubit 2. As the number of layers is further increased, the number of accessible frequencies is reduced in qubit 4 and then in qubits 2 and 3.
These findings are consistent with \cite{casas_multidimensional_2023}, where it is suggested that the number of adjustable degrees of freedom will saturate at two circuit layers. In comparison, the best accuracy is achieved at 2 layers for the BreastMNIST dataset, followed with a drop in accuracy at 3 layers, which aligns again with these results. On the OrganAMNIST dataset, the best performance is attained at 3 layers but with only a marginal improvement observed from the second to the third layer.


For $R_{Y}$ and the higher-order encoding, increasing the number of data-reuploading layers causes the variance of the Fourier coefficients to increase and their values to become more mixed (real and imaginary). For both encodings, this increase is mostly noticeable when transitioning from one to two circuit layers, and becomes less visible when increasing further. This is an indication of a saturation of adaptable degrees of freedom at two layers, which was expected for $R_{Y}$, but was expected at 4 layers for the higher-order encoding. Regarding performance, the best performance with $R_{Y}$ occurs at 4 and 2 layers on BreastMNIST and OrganAMNIST, respectively. The findings on OrganAMNIST are in line with both the observed distribution of Fourier coefficients and the previous work on the saturation of adjustable degrees of freedom. However, the result on BreastMNIST diverges, likely due to the limited size of the dataset. For the higher-order encoding, the best outcomes are achieved at 3 and 4 layers, indicating the relevance of the analysis in \cite{casas_multidimensional_2023}.



\section{Conclusion}
We analyze the performance of various hybrid quantum-classical convolutional neural networks composed of different encodings on two medical imaging tasks, namely breast cancer detection on ultrasound images and organ classification with CT scans. We first observe that the encoding used -- be it $R_{X}$, $R_{Y}$ angle encoding, higher-order, or amplitude encoding -- leads to significantly different performances, highlighting the critical significance of encoding selection when implementing quantum variants of classical algorithms.
We additionally examine the effects of input scaling and data-reuploading and show that a well-chosen parametrization of these two elements can also lead to a substantial gain in performance on a chosen dataset.
These findings point to the need for continued investigation into QCCNNs, considering their compatibility with current NISQ hardware thanks to the minimal qubit demands and shallow quantum circuit depths involved.
When attempting to explain these performances using quantum metrics such as the expressibility, the entanglement capability, and the normalized effective dimension of the circuit, a clear correlation is yet to be observed. On the other hand, a discernible trend begins to emerge upon closer examination of the Fourier coefficients generated by the quantum circuits, where a high number of non-null frequency as well as a wide-spread distribution of coefficients across both real and imaginary axes is linked to a good performance. Moreover, one can observe changes in the distribution of coefficients, thus indicating potential saturation effects as additional reuploading layers are incorporated. Still, additional work is needed to understand how these findings can effectively inform encoding selection based on the dataset employed. Furthermore, future work should also focus on investigating the impact of noise on these architecture, via simulation packages, but also by running the results on the hardware to comprehensively assess its impact.
\label{ccl}

\section{Acknowlegements}

The project/research is supported by the Bavarian Ministry of Economic Affairs, Regional Development and Energy with funds from the Hightech Agenda Bayern.
\bibliographystyle{IEEEtran}
\bibliography{IEEEabrv, bibliography}

\end{document}